\documentclass[twocolumn,prl,aps,showpacs,superscriptaddress,amsmath,amssymb]{revtex4}

\usepackage{graphicx}
\usepackage{dcolumn}
\usepackage{bm}

\newcommand{\bfr}{{\mathbf{r}}}
\newcommand{\ee}{{\mathrm{e}}}
\newcommand{\calG}{{\mathcal{G}}}
\newcommand{\dd}{{\mathrm{d}}}
\newcommand{\PsiGML}{{\Psi_{\mathrm{GML}}}}

\newcommand{\barE}{\overline{E}}

\newcommand{\barphi}{\overline{\varphi}}

\begin{document}

\preprint{APS/123-QED}

\title{The many-body Green function of degenerate systems}%

\author{Christian Brouder}
\affiliation{%
Institut de Min\'eralogie et de Physique des Milieux Condens\'es,
CNRS UMR 7590, Universit\'es Paris 6 et 7, IPGP, 140 rue de Lourmel,
75015 Paris, France.
}%
\author{Gianluca Panati}
\affiliation{%
Dipartimento di Matematica, Universit\`a di Roma La Sapienza, Roma,
Italy.
}%
\author{Gabriel Stoltz}
\affiliation{%
Universit\'e Paris Est, CERMICS, Projet MICMAC ENPC - INRIA, \\
  6 \& 8 Av. Pascal, 77455 Marne-la-Vall\'ee Cedex 2, France.
}%

\date{\today}

\begin{abstract}
A rigorous non perturbative adiabatic approximation of the
evolution operator in the many-body
physics of degenerate systems is derived. This approximation is used
to solve the long-standing problem of the choice of the 
initial states of $H_0$ leading to eigenstates of $H_0+V$
for degenerate systems.
These initial states are eigenstates of 
$P_0 V P_0$, where $P_0$ is the projection onto a degenerate
eigenspace of $H_0$.
This result is used to give the proper definition of the
Green function, the statistical Green function and the
non-equilibrium Green function of degenerate systems.
The convergence of these Green functions is established.
\end{abstract}

\maketitle

%

Non-perturbative Green function methods, such as the 
GW approximation~\cite{GW} or the Bethe-Salpeter
equation~\cite{Onida,Lawler}, have brought remarkable
progress in the calculation of the electronic 
structure and dielectric response of semiconductors.
The extension of these methods to transition metals
systems faces a serious difficulty: the standard Green function can
only be defined when the initial
state $|0\rangle$ of the system without interaction
is a single Slater determinant. In physical terms,
each single-particle orbital or Bloch state has
to be either occupied or unoccupied at zero temperature.
However, the physics of transition metals often contradicts
this requirement. For the example of a V$^{3+}$ ion in an
octahedral environment, we do not know a priori how the 
two $3d$ electrons are distributed over the six degenerate
$t_{2g}$ orbitals (with up and down spins).

More generally, for a system described by a Hamiltonian
$H=H_0+V$ where the ground state of $H_0$ is degenerate,
we need to determine the \emph{parent states},
i.e. the initial states of $H_0$ that evolve into
eigenstates of $H$ by adiabatically switching the interaction.

Degenerate systems being ubiquitous in quantum physics,
this long-standing problem has been discussed in
chemical physics~\cite{TolmachevI,Banerjee}, 
nuclear physics~\cite{Kuo,Coraggio}, 
atomic physics~\cite{Braun,Lindgren3} and 
solid state physics~\cite{Esterling}. 
Esterling and Lange~\cite{Esterling} summarized the situation
as follows:
``Since $H_0$ has degenerate ground states, the choice of the state 
$|0\rangle$ must be made with care, and this may be considered the 
key to the problem.''
This question is also crucial in many-body physics because
the Green function of a degenerate system
has to be defined from a parent state.

In the present paper, we give a simple method to explicitly
determine the parent states and to define the Green
function of degenerate systems.
Through a non-perturbative analysis of the
evolution operator of a degenerate system, we determine
the exact form of its singularities. This enables us to derive:
(i) an easy and explicit method to determine the parent states;
(ii) a non-perturbative proof that the Gell-Mann and Low formula 
generally converges only for these parent states;
(iii) the formula for the Green function of degenerate systems;
(iv) the validity of the so-called
  \emph{statistical} Green function;
(v) the singularity structure of the non-equilibrium
  Green function of degenerate systems.

\emph{Adiabatic switching}.
Many-body theory~\cite{Gross,Fetter} is usually based on the adiabatic 
switching of the
interaction, i.e. the transformation of the time-independent
Hamiltonian $H=H_0+V$ into the time-dependent one
$H_0+ \ee^{-\varepsilon |t|} V$. Adiabatic switching 
turns the non degenerate ground state $|0\rangle$ of $H_0$
into an eigenstate $|\PsiGML\rangle$ of $H$ 
first proposed by Gell-Mann and Low~\cite{GellMann} in 1951
\begin{eqnarray}
|\PsiGML\rangle = 
\lim_{\varepsilon\to0} \frac{U_\varepsilon(0,-\infty)|0\rangle}
                         {\langle 0 |U_\varepsilon(0,-\infty)|0\rangle},
\label{GMLdef}
\end{eqnarray}
where the evolution operator $U_\varepsilon(t,t')$ is the solution of
\begin{eqnarray*}
i\frac{\partial U_\varepsilon(t,t')}{\partial t} = 
\ee^{iH_0 t} \ee^{-\varepsilon |t|} V \ee^{-iH_0 t} U_\varepsilon(t,t'),
\end{eqnarray*}
with the initial condition $U_\varepsilon(t',t')=1$.
The wavefunction $|\PsiGML\rangle$ is then used to build the
Green function of the system \cite{Fetter,Gross}.
However, Gell-Mann and Low did not prove that the limit of
eq.~(\ref{GMLdef}) exists~\cite{Gross}.
The convergence of 
$|\PsiGML\rangle$ for nondegenerate systems was first established
by Nenciu and Rasche in 1989~\cite{Nenciu}.

For a
degenerate ground state $|0\rangle$ of $H_0$, the Gell-Mann and Low formula 
generally fails to converge when $\varepsilon\to0$, as can
be seen even for a trivial two-level system~\cite{BSP}.
In the following, we use recent advances 
in the mathematical analysis of the adiabatic approximation
(see \cite{Teufel} for a review)
to extend the Gell-Mann and Low formula to degenerate systems.

\begin{figure}
\includegraphics[width=8.0cm]{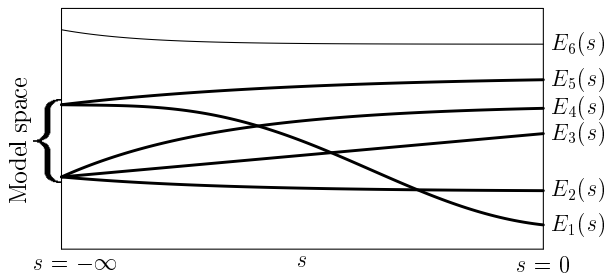}
\caption{Exemple of allowed eigenvalue pattern\label{figE}}
\end{figure}

\emph{Adiabatic approximation.}
In this section, we set up the notation and give the
theorem that enables us to calculate the parent states
and to define the Green functions.
We consider $t\le0$ and we rewrite the time-dependent Hamiltonian
as $H(s)=H_0+\ee^s V$, where $s=\varepsilon t$ is the so-called
\emph{slow variable}.
The eigenvalues of $H(s)$ are denoted by $E_j(s)$
and its eigenprojectors by $P_j(s)$.
We recall that, if the eigenvalue $E_j(s)$ is $n_j$-fold degenerate,
the eigenprojector is 
$P_j(s)=\sum_{k=1}^{n_j} |\varphi_{jk}(s)\rangle \langle \varphi_{jk}(s)|$,
where $\{|\varphi_{jk}(s)\rangle\}$ is a set of $n_j$ orthonormal
eigenstates of $H(s)$ for the eigenvalue $E_j(s)$.
For notational convenience, we denote $P_j(-\infty)$
by $P_j^0$ in the rest of the paper.  
The \emph{model space} $M$ is the vector space generated by
the eigenstates corresponding to $N_0$ eigenvalues of 
$H_0=H(-\infty)$. Each eigenvalue of $H_0$ can be degenerate
and is possibly split by the perturbation $V$, so that the
$N_0$ eigenvalues of $H_0$ become $N$ eigenvalues 
$E_1(s), \dots, E_N(s)$ of $H(s)$, with $N\ge N_0$.
Each $E_j(s)$ can be degenerate 
and the eigenvalues are allowed to cross
(fig.~\ref{figE}).
For an octahedral $V^{3+}$ ion, we have $N_0=1$
with degeneracy 15,
and there are $N=4$ interacting states:
${}^1A_{1g}$,
${}^1E_{g}$,
${}^1T_{2g}$ and
${}^3T_{1g}$, with degeneracy
$n_j$=1, 2, 3 and 9, respectively.

A key tool of our approach 
is $A(s,s_0)$,
the \emph{rotating frame
operator}~\cite{Kato,Messiah-GB},
that relates the eigenstates at $s_0$ and $s$:
$A(s,s_0)|\phi_{jk}(s_0)\rangle=|\phi_{jk}(s)\rangle$,
so that
\begin{eqnarray}
A(s,s_0)P_j(s_0) &=& P_j(s) A(s,s_0).
\label{intertwin}
\end{eqnarray}
Using standard technical assumptions~\cite{relatively-bounded},
we recently obtained~\cite{BPS} a rigorous approximation of the evolution
operator projected on each eigenspace:
\begin{eqnarray}
   U_\varepsilon(0,-\infty) P_j^0
  \simeq 
\ee^{i\theta_j/\varepsilon} A(0,-\infty)P_j^0,
\label{UA}
\end{eqnarray}
where
$\theta_j = -\int_{-\infty}^0 (E_j(s)-E_j(-\infty)) \dd s$.
In particular, the divergences of the evolution operator are
entirely described by the factor $\ee^{i\theta_j/\varepsilon}$.

\emph{Construction of the parent states.}
The parents states are the eigenstates
$|\phi\rangle$ of $H_0$ such that
$U_\varepsilon(0,-\infty)|\phi\rangle$
tends to an eigenstate $|\Psi\rangle$ of $H$, up to a (divergent)
phase. Therefore, the parent states are naturally
defined in terms of $U_\varepsilon(-\infty,0)|\Psi\rangle$
and it seems that the 
interacting states $|\Psi\rangle$ are needed 
to define the parent 
states~\cite{TolmachevI,KuoOsnes,Lindgren3}.
We now show that the parent states
have a more simple and explicit definition
as eigenstates
of $P_j^0$ and we explain how $P_j^0$ can be calculated 
by standard time-independent perturbation theory.

For notational convenience, we
denote $\ee^s$ by $\lambda$ and the
eigenvalues and eigenprojections are written
in terms of $\lambda$. 
We denote by 
$\barE_j(\lambda)$ and $|\barphi_{jk}(\lambda)\rangle$
the eigenvalues and eigenstates
of $H_0+\lambda V$ (so that
$\barE_j(\lambda)=E_j(s)$).
They can be expanded as~\cite{BPS}
\begin{eqnarray*}
\barE_j(\lambda) = \sum_{n=0}^\infty \lambda^n E_j^n,\quad
|\barphi_{jk}(\lambda)\rangle =
\sum_{n=0}^\infty \lambda^n |\varphi_{jk}^n\rangle,
\end{eqnarray*}
with the normalization
$\langle \varphi_{jk}^0 | \barphi_{jk}(\lambda)\rangle =1$.
The eigenstates $|\barphi_{jk}(\lambda)\rangle$ are 
assumed orthonormal only at $\lambda=0$, where 
$P_j^0 = \sum_{k=1}^{n_j}
|\varphi_{jk}^0\rangle \langle \varphi_{jk}^0|$.

The time-independent Schr\"odinger equation
\begin{eqnarray*}
(H_0+\lambda V) |\varphi_{jk}(\lambda)\rangle &=&
E_j(\lambda)|\varphi_{jk}(\lambda)\rangle,
\end{eqnarray*}
gives, to order 0,
$H_0 |\varphi_{jk}^0\rangle = E_j^0 |\varphi_{jk}^0\rangle$, 
so that 
$|\varphi_{jk}^0\rangle$ is an eigenstate of $H_0$ with energy $E^0_j$.
We assume that $E^0_j$ is one of the
$N_0$ eigenvalues of the model space, so that
$|\varphi_{jk}^0\rangle$ belongs to the model space.
However, the degeneracy of $E^0_j$ as an eigenvalue
of $H_0$ is generally larger than the degeneracy of
$E_j(\lambda)$ and we need more information
to determine the $n_j$ states $|\varphi_{jk}^0\rangle$.
The Schr\"odinger equation to order $\lambda$ gives us
$(H_0 -E^0_j) |\varphi_{jk}^1\rangle = (E_j^1 -V) |\varphi_{jk}^0\rangle$.
This equation can only 
have a solution if $\langle \psi_m^0 |(E_j^1 - V) |\varphi_{jk}^0\rangle=0$,
where $\{|\psi_m^0\rangle\}$ is a complete set of eigenstates of $H_0$
with energy $E^0_j$. Therefore, 
the initial states $|\varphi_{jk}^0\rangle$ are eigenstates of $H_0$
with energy $E^0_j$ and eigenstates of $P_{E^0_j} V P_{E^0_j}$ with eigenvalue
$E^1_j$, where $P_{E_j^0}$ is the projection onto the 
eigenspace of $H_0$ with eigenvalue $E^0_j$.
In general, the degeneracy is split at this order,
in the sense that there are only $n_j$ states that are 
simultaneously eigenstates of $H_0$ with energy $E^0_j$ and 
of $P_{E^0_j} V P_{E^0_j}$ with eigenvalue $E^1_j$. 
Otherwise, for instance when $P_{E^0_j} V P_{E^0_j}$
is zero by symmetry, the equations coming from higher
powers of $\lambda$ must be taken into account to determine
$|\varphi_{jk}^0\rangle$. 
In that case, the second order is usually enough~\cite{Hirschfelder-69},
but methods have been developed to treat
any order~\cite{Tsaune}.

We generally have no a priori knowledge of $n_j$
and $E^1_j$. However, we can calculate all the
eigenstates of $H_0$ and, for each energy $E_j^0$,
we can diagonalize $P_{E^0_j} V P_{E^0_j}$.
Then, each state must be examined to
see if it cannot be further split by higher order terms.
When degeneracy is due to the symmetry of the Hamiltonian
$H(s)$, this can be deduced from the dimension of 
the irreducible representations to which the states belong.
The computational effort required to construct $|\varphi^0_{jk}\rangle$
is small because it is an eigenvalue problem in a vector
space whose dimension is the degeneracy of $E_j^0$, which
is small in applications.
From the states $|\varphi^0_{jk}\rangle$
we build the projector $P_j^0$ and 
we define a \emph{parent state} as a 
state $|\phi\rangle$ such that, for some $j$,
\begin{eqnarray}
P_j^0 |\phi\rangle = |\phi\rangle.
\label{projection}
\end{eqnarray}
In practice, the parent state is one of the 
$|\varphi^0_{jk}\rangle$.

\emph{Generalized Gell-Mann and Low wavefunction.}
We show that the parent states previously defined
lead to convergent Gell-Mann and Low wavefunctions.
For a parent state $|\phi_j\rangle$ such that
$P^0_j|\phi_j\rangle=|\phi_j\rangle$,
eq.~(\ref{UA}) enables us to write
\begin{eqnarray*}
U_\varepsilon(0,-\infty) |\phi_{j}\rangle &=&
U_\varepsilon(0,-\infty) P^0_j|\phi_{j}\rangle 
\\
& \simeq &
\ee^{i\theta_j/\varepsilon} A(0,-\infty)P_j^0 |\phi_{j}\rangle 
\\
& \simeq &
\ee^{i\theta_j/\varepsilon} A(0,-\infty)|\phi_{j}\rangle.
\end{eqnarray*}
Therefore, the following limit exists:
\begin{eqnarray}
|\PsiGML\rangle &=& 
\lim_{\varepsilon\to0} \frac{U_\varepsilon(0,-\infty)|\phi_{j}\rangle}
   {\langle \phi_{j} |U_\varepsilon(0,-\infty)|\phi_{j}\rangle}
\label{PsiGML}\\&=&
\frac{A(0,-\infty)|\phi_{j}\rangle}
   {\langle \phi_{j} |A(0,-\infty)|\phi_{j}\rangle}.
\nonumber
\end{eqnarray}
The Gell-Mann and Low wavefunction $|\PsiGML\rangle $ is indeed
an eigenstate of $H_0+V$ with energy $E_j(0)$
because $P_j(0)|\PsiGML\rangle = |\PsiGML\rangle $.
To show this, we use eq.~(\ref{intertwin}):
\begin{eqnarray*}
P_j(0) A(0,-\infty)|\phi_{j}\rangle
&=&
A(0,-\infty) P_j^0|\phi_{j}\rangle
\\&=&
 A(0,-\infty) |\phi_{j}\rangle.
\end{eqnarray*}

In practice, we are interested in the Gell-Mann and
Low wavefunction that is the ground state of
$H_0+V$. How should we choose the initial state
$|\phi_{j}\rangle$ for this to happen?
In the non degenerate case, it is often assumed
that the ground state of $H_0$ leads to
the ground state of $H_0+V$. 
When degeneracy is due to the presence of symmetry,
band crossing can occur and one should try 
the $|\phi_{j}\rangle$ corresponding to the lowest energy
$E^0_j$ for each irreducible representation. 
A typical example of band crossing in the
presence of symmetry is given by Tanabe-Sugano
diagrams of the multiplet theory~\cite{TanabeSugano}.
For a small crystal field, the ground state has
the highest spin value (Hund's rule), but as the
crystal field parameter increases, a low spin state
can become the ground state.

\emph{Green functions.}
The expression for the Green function is usually
derived under the assumption that the ground state
of $H_0$ is non degenerate~\cite{Fetter,Gross}.
Our results enable us to 
determine how this expression is extended to the
case of degenerate systems.
Now we formally extend our previous results
to Fock space.

To follow the usual argument~\cite{Gross}, we 
repeat the calculation by starting from a positively
infinite time. 
In terms of the slow variable
$s=-\varepsilon|t|$, the switching function 
$\ee^{-\varepsilon|t|}$ is the same
for positive and negative times. As a result, the
rotating frame operator is the same but the divergent
phase changes sign:
$U_\varepsilon(0,+\infty) |\phi_{j}\rangle
\simeq 
\ee^{-i\theta_j/\varepsilon} A(0,-\infty) |\phi_{j}\rangle$. 
Therefore,
\begin{eqnarray*}
\lim_{\varepsilon\to0} \frac{U_\varepsilon(0,+\infty)|\phi_{j}\rangle}
   {\langle \phi_{j} |U_\varepsilon(0,+\infty)|\phi_{j}\rangle}
=
\frac{A(0,-\infty)|\phi_{j}\rangle}
   {\langle \phi_{j} |A(0,-\infty)|\phi_{j}\rangle}.
\end{eqnarray*}
In other words, the Gell-Mann and Low wavefunctions
obtained from positive and negative infinite times are
equal. This non-trivial result is due to
the fact that the switching function
$f(t)=\ee^{-\varepsilon|t|}$ is even.

The two-point Green function is defined by~\cite{Gross}
\begin{eqnarray*}
G(x,y) &=& \frac{\langle\PsiGML| O_H |\PsiGML\rangle}
{\langle\PsiGML|\PsiGML\rangle},
\end{eqnarray*}
where $x=(\bfr,t)$, $y=(\bfr',t')$,
 $O_H= T\big(\psi_H(x)\psi_H^\dagger(y)\big)$ is the
time-ordered product of fields in the Heisenberg
picture and $|\PsiGML\rangle$ is defined by
eq.~(\ref{PsiGML}).
Standard manipulations~\cite{Gross} transform it into 
\begin{eqnarray}
G(x,y) &=& 
\lim_{\varepsilon\to 0}
\frac{\langle\phi_{j}|X_\varepsilon|\phi_{j}\rangle}
{\langle\phi_{j}|U_\varepsilon(+\infty,-\infty)|\phi_{j}\rangle},
\label{Green}
\end{eqnarray}
where
$X_\varepsilon = 
U_\varepsilon(+\infty,t)
 \psi(x)U_\varepsilon(t,t')\psi^\dagger(y)U_\varepsilon(t',-\infty)$
and
$X_\varepsilon = 
- U_\varepsilon(+\infty,t')
 \psi^\dagger(y)U_\varepsilon(t',t)\psi(x)U_\varepsilon(t,-\infty)$
if $t>t'$ and $t<t'$, respectively.

The expression for the Green function
generally converges only when the initial state is
a parent state. Indeed,
consider a state $|\phi\rangle$ in the model space and write it
as $|\phi\rangle=\sum_j |\phi_j\rangle$, where
$|\phi_j\rangle = P_j^0 |\phi\rangle$.
Thus,
\begin{eqnarray*}
U_\varepsilon(0,\pm\infty)|\phi\rangle \simeq
\sum_j \ee^{\mp i \theta_j/\varepsilon} A(0,-\infty)|\phi_j\rangle.
\end{eqnarray*}
If there is more than one $j$ in the sum, 
the phases $\theta_j$ are generally different
(in the absence of eigenvalue crossing, they can be shown
to be different). Therefore, the phase factors in the
numerator and denominator of eq.~(\ref{Green}) do not cancel
and the expression has no limit for $\varepsilon\to0$.

\emph{Statistical Green function}.
The Green function of the previous section has a non-ambiguous
meaning when $|\phi_j\rangle$ is the parent state of a non-degenerate
interacting state. However, when the interacting state itself
is degenerate, there
is no reason to choose any particular parent state.
To solve that problem, Layzer~\cite{Layzer} defined the
\emph{statistical Green function} as
an equal-weight average over the degenerate states. Such a statistical 
Green function was advocated, for instance, by Alon and
Cederbaum~\cite{Alon}. The statistical Green function can
preserve the symmetry of the system: in the example
of a spherically symmetric Hamiltonian, the Green function
obtained from any state 
$|\ell m\rangle$ with $\ell\not=0$ gives non spherically
symmetric charge density, whereas the statistical Green function
obtained from the mixed state
$\sum_m |\ell m\rangle (2\ell+1)^{-1} \langle \ell m|$
gives a spherical charge density~\cite{Lieb}.
We are now able to prove that the statistical Green function
has a well-defined limit when $\varepsilon\to 0$.

To define the statistical Green function
of a degenerate interacting system with energy $E_j(0)$,
we use the density matrix
$\rho=(1/n_j) \sum_{k=1}^{n_j}
|\varphi_{jk}^0\rangle\langle\varphi_{jk}^0|$,
where the states $|\varphi_{jk}^0\rangle$ are those used to calculate
$P_j^0$. We assume that the degeneracy of 
$E_j(0)$ is $n_j$.  Then,
\begin{eqnarray}
G(x,y) &=& 
\lim_{\varepsilon\to 0}
\frac{\mathrm{Tr}(\rho X_\varepsilon)}
{\mathrm{Tr}\big(\rho U_\varepsilon(+\infty,-\infty)\big)}.
\label{statGreen}
\end{eqnarray}
If we put $|\phi_j\rangle=|\varphi_{jk}^0\rangle$
in eq.~(\ref{Green}), the divergences of the numerator and
denominator are $\ee^{2i\theta_j/\varepsilon}$, 
which \emph{does not depend} on $k$.
Therefore, the divergent phases of the numerator and
denominator of eq.~(\ref{statGreen}) are equal, and the statistical
Green function is well defined.
For our octahedral $V^{3+}$ ion, the Green function is defined
with the density matrix built from the 
nine degenerate states with 
${}^3T_{1g}$ symmetry if the ion is high spin.

\emph{Non-equilibrium Green function}.
In the study of non-equilibrium systems, it is often convenient
to run the evolution operator over a closed time path instead of
taking the limit $t\to+\infty$. 
In this so-called Keldysh approach, the Green function
$\calG(x,y)$ 
is calculated by a formula involving no denominator~\cite{Rammer-07}:
it is the limit for $\varepsilon\to 0$ of
$\calG_\varepsilon = \langle \phi| U_\varepsilon(-\infty,0) O_H
   U_\varepsilon(0,-\infty) |\phi\rangle$.
As for the standard Green function, this expression generally 
converges for degenerate systems only when $|\phi\rangle$ is a parent
state. To see this, we expand again a state of the model
space over parent states: $|\phi\rangle=\sum_j|\phi_j\rangle$.
Then,
\begin{eqnarray*}
\calG_\varepsilon \simeq
  \sum_{ij} \ee^{i(\theta_j-\theta_i)/\varepsilon}
\langle \phi_i| A(-\infty,0) O_H A(0,-\infty) |\phi_j\rangle.
\end{eqnarray*}
This expression converges for $\varepsilon\to0$
when there is a single phase,
i.e. when $|\phi\rangle$ is a parent state. 
Otherwise, the limit generally does not exist.


\emph{Conclusion}.
The determination of the parent states and the proof of convergence
break the last
deadlocks in the determination of the Green function of
degenerate systems.
The main difference with the non-degenerate case is the fact that,
for many degenerate systems, the parent state is not a single 
Slater determinant. 
To see this, consider a Hamiltonian where
$H_0$ is the restricted Hartree-Fock Hamiltonian of a
$3d^n$ transition metal ion and $V$ is the sum of the remaining 
atomic Coulomb interaction and of an effective potential
representing the influence of the surrounding atoms.
Then, the parent states are exactly the eigenstates of 
the crystal-field Hamiltonian and they are generally
not single Slater determinants.
In that case, the structure of the Green function is
more complex because of the so-called 
\emph{initial correlations}~\cite{Hall} coming from
the matrix elements between the different Slater determinants.
Perturbative~\cite{Hall} and 
nonperturbative~\cite{BrouderBSL} methods have
been developed to tackle initial correlations.
Finally, our approach also gives a non-perturbative proof of the
convergence of the effective Hamiltonian~\cite{Weber-01}.


\begin{thebibliography}{31}
\bibitem[{rel()}]{relatively-bounded}
\bibinfo{note}{Our assumptions hold for a Hamiltonian describing the many-body
  Coulomb interaction of electrons and nuclei~\cite{Kato}, provided there is a
  gap between the model space and the rest of the spectrum.}
\end{thebibliography}

\end{document}